\documentclass [12pt] {article}
\usepackage{graphicx}
\usepackage{epsfig}
\def\lsim{\raise0.3ex\hbox{$<$\kern-0.75em\raise-1.1ex\hbox{$\sim$}}}
\def\gsim{\raise0.3ex\hbox{$>$\kern-0.75em\raise-1.1ex\hbox{$\sim$}}}

\begin {document}

\begin{center}
{\Large {\bf $\Lambda$-BARYON PRODUCTION  IN}}\\
\vskip 0.3cm
{\bf {\Large {$\pi^{\pm}$N INTERACTIONS}}} \\

\vskip 1.5 truecm
{ G. H. Arakelyan$^1$, C. Merino$^2$, and Yu. M. Shabelski$^3$}\\
\vskip 0.5 truecm
Yerevan Physics Institute, Armenia\\
\end{center}

\vskip 1.5 truecm

\begin{center}
{\bf ABSTRACT}
\end{center}

The process of $\Lambda$-baryon production in $\pi p$ collisions is considered.
The  contribution of  the string-junction  mechanism to the strange baryon production in
meson-baryon scattering is analysed.  The results of numerical
calculations in the framework of the Quark-Gluon String model are in
reasonable agreement with the data.

\vskip 1.5 truecm

\noindent $^1$Permanent address: Yerevan Physics Institute,  Armenia\\
E-mail: argev@mail.yerphi.am
\vskip 0.3 truecm
\noindent $^2$Permanent address:  Departamento de F\'\i sica de Part\'\i culas, Facultade de F\'\i sica, and
Instituto Galego de Altas Enerx\'\i as (IGAE),
Universidade de  Santiago de Compostela, Galicia, Spain \\
E-mail: merino@fpaxp1.usc.es

\vskip 0.3 truecm
\noindent $^3$Permanent address: Petersburg Nuclear Physics Institute,
Gatchina, St.Petersburg, Russia\\
E-mail: shabelsk@thd.pnpi.spb.ru

\newpage
\pagestyle{plain}
\noindent{\bf 1. INTRODUCTION}
\vskip 0.5 truecm

The Quark--Gluon String Model (QGSM) is based on the Dual Topological
Unitarization (DTU) and it describes quite reasonably many
features of high energy production processes, including the inclusive spectra of
different secondary hadrons, their multiplicities, KNO--distributions, etc., both
in hadron--nucleon and hadron--nucleus collisions \cite{KTM,2r,KTMS,Sh}.
High energy interactions are considered as  taking place via the exchange
of one or several pomerons, and all elastic and inelastic processes
result from cutting through or between  those exchanged pomerons \cite{AGK,TM}. The
possibility of different  numbers of pomerons  to be exchanged  introduces
absorptive corrections to the  cross-sections which are in
agreement with the experimental data on production of hadrons consisting
of light quarks. Inclusive spectra of hadrons are related to the
corresponding fragmentation functions of quarks and diquarks, which
are constructed using the reggeon counting rules \cite{Kai}.

In  previous papers \cite{ACKS, ShBopp}  one has discussed the processes connected with the transfer
of baryon charge over long rapidity distances. In the string models
baryons are considered as configurations consisting of three strings
attached to three valence quarks and connected in  one point
called ``string junction" (SJ)  \cite{IOT,RV}. Thus the string-junction has a nonperturbative origin in QCD.

It is very important to understand the role of the SJ in
the dynamics of high-energy hadronic interactions, in particular in
processes implying baryon number transfer \cite{14ra,14rb,Khar}.  Significant results
on this question were obtained in \cite{ACKS, ShBopp}. The difference between \cite{ACKS} and \cite{ ShBopp}
is only in the values of  Regge intercept of  SJ.

In  the present paper we  extend our study   to the case of strange baryon production in $\pi N$  interactions.

In  references \cite{ACKS, ShBopp} the SJ mechanism
was mainly analyzed for proton production in $\pi p$ and $pp$ collisions, and strange baryon
production in $pp$ interaction. For the $\Lambda$ production on pion beam were only considered
the data on asymmetry.

 Now we consider the data on spectra and  new appeared data on  asymmetry of $\Lambda$
 production on  $\pi$-beam,  not analysed in \cite{ACKS, ShBopp}. For consideration
 we used both parametrisations of diquark fragmentation functions to strange baryons and Regge
 trajectory intercepts, as in  \cite{ACKS, ShBopp} (see sections 2 and 3 below).

In this paper we present a description of all  available data  and
extract information on the properties of the SJ dynamics.

\vskip 0.9 truecm \noindent{\bf 2. INCLUSIVE SPECTRA OF SECONDARY HADRONS IN
QGSM} \vskip 0.5 truecm

As  it is thoroughly known, the exchange of one or several pomerons is one basic feature of
high energy hadron--nucleon and hadron--nucleus interactions
 in the frame of QGSM and DPM. Each pomeron corresponds to a cylindrical  diagram,
and thus, when cutting a pomeron two showers of secondaries
are produced . The inclusive spectrum of secondaries is
determined by the convolution of diquark, valence, and sea quark
 distribution functions in the incident particle, $u(x,n)$, and the fragmentation
functions of quarks and diquarks into secondary hadrons, $G(z)$.

The diquark and quark distribution functions depend on the number $n$ of cut
pomerons in  a given diagram. In the following we  will use the formalism of
QGSM. In the case of a nucleon target the inclusive spectrum of a secondary
hadron $h$ has the form \cite{KTM}:

\begin{equation}
\label{t0}
\frac{{x_E}}{\sigma_{inel}}
\frac{d\sigma}{dx} =\sum_{n=1}^{\infty}w_{n}\phi_{n}^{h}(x),
\end{equation}
where $x$ is the Feynman variable, and $x_E=2E/\sqrt s$.

The probability for a process to have $n$ cutted pomerons, $w_{n}$, has
been extensively described using the quasi\-ei\-ko\-nal approximation
(see, for instance, \cite{KTM,TM,ACKS}).

The function $\phi_{n}^{h}(x)$ represents the contribution of diagrams
with $n$ cut pomerons. In the case of a meson beam it has the form:

\begin{equation}
\label{t1}
\phi_{\pi p}^{h}(x) = f_{\bar{q}}^{h}(x_{+},n)\cdot f_{q}^{h}(x_{-},n) +
f_{q}^{h}(x_{+},n)\cdot f_{qq}^{h}(x_{-},n) +
2(n-1)f_{s}^{h}(x_{+},n)\cdot f_{s}^{h}(x_{-},n),
\end{equation}
with

\begin{equation}
\label{t2}
x_{\pm} = \frac{1}{2}\Big[\sqrt{4m_{T}^{2}/s+x^{2}}\pm{x}\Big],
\end{equation}
where $f_{qq}$, $f_{q}$, and $f_{s}$ correspond to the contributions of
diquarks, valence quarks and sea quarks, respectively, and they are
determined by the convolution of the diquark and quark distributions with the
fragmentation functions, e.g.,

\begin{equation}
\label{t3}
f_{q}^{h}(x_{+},n) = \int_{x_{+}}^{1} u_{q}(x_{1},n)\cdot G_{q}^{h}(x_{+}/x_{1})
dx_{1}.\ \
\end{equation}

The diquark and quark distributions, as well as the fragmentation
functions, are determined from Regge intercepts \cite{KTMS,Sh} (for the
mathematical form of distribution functions see, for instance, \cite{ACKS}).

In the present paper we used the expression of the quark
fragmentation functions into $\Lambda$ given in \cite{ACKS,ShBopp}:

\begin{equation}
\label{t4}
G_{u}^{\Lambda} = G_d^{\Lambda} = {a_{\bar{\Lambda}}}
(1-z)^{\lambda + \alpha_R - 2 \alpha_B+\Delta \alpha }\cdot (1+a_{1}z^{2}) \;\;, G_u^{\bar{\Lambda}} = G_{\bar{d}}^{\Lambda}= (1-z)\cdot G_d^{\Lambda},
\end{equation}
with

\begin{equation}
\label{t5}
\Delta \alpha = \alpha_{\rho} - \alpha_{\phi} = 1/2 \;\; ,~~~~
\lambda=2\alpha^{\prime} < p_{t}^2>=0.5 \;.
\end{equation}

Diquark fragmentation functions have more complicated forms. They contain
two contributions. The first one corresponds to the central
production of a $B\bar{B}$ pair, and has the form (see \cite{ACKS}):

\begin{equation}
\label{t5a}
G_{uu}^{\Lambda} = G_{ud}^{\Lambda} = G_{uu}^{\bar{\Lambda}}
= G_{ud}^{\bar{\Lambda}} = a_{\bar{\Lambda}}
(1-z)^{\lambda-\alpha_R + 4(1-\alpha_B)+{\Delta \alpha}}. \;\;
\end{equation}

The second contribution is connected with the direct fragmentation of
the initial baryon into the secondary one with conservation of the
string junction. As discussed in \cite{ACKS, ShBopp}, three different
types of such contributions exist, the secondary baryon consisting of
the SJ together with: a) two valence and one sea quarks, b) one valence
and two sea quarks, or c) three sea quarks (see Fig.~2 in
\cite{ACKS,ShBopp}). The fraction of the incident baryon energy carried
by the secondary baryon decreases from cases a) to c), whereas the mean
rapidity gap between the incident and secondary baryon increases.

The probability to find a comparatively slow SJ in the  case c)
can be estimated from the data on $\bar{p}p$ annihilation  into
mesons  (see  \cite{ACKS}), where this probability, only experimentally known at
low energies, turns out to be proportional to
$s^{\alpha_{SJ}-1}$. However, it has been also argued \cite{14r} that the annihilation cross section
contains a small component which is independent of $s$, making $\alpha_{SJ}\sim1$.

The contribution of the  term c) was analyzed  in \cite{ACKS, ShBopp}, where its magnitude was taken
proportional to a coefficient that we will denote by $\epsilon$.

 All the contributions are determined by equations similar to (\ref{t1}-\ref{t3}),
with the corresponding fragmentation functions.

As it was  noted above, we compare our calculations using the both  
parametrizations of fragmentation functions of diquarks into strange 
baryons and corresponding SJ trajectory intercepts~$\alpha_{SJ}$\cite{ACKS, ShBopp}.

In references \cite{ACKS,ShBopp} were used the following expressions:

\begin{equation}
\label{t7}
 G_{ud}^{\Lambda} = a_N z^{\beta}
\Big[v_0\varepsilon (1-z)^2 + v_q z^{2 - \beta} (1-z) +
v_{qq}z^{2.5 - \beta}\Big](1-z)^{\Delta \alpha}\; , \;
G_{uu}^{\Lambda} =(1-z)G_{ud}^{\Lambda}. \;
\end{equation}
with $\beta = 1 - \alpha_{SJ}$.

The factor $z^{\beta}$ is actually $z^{1-\alpha_{SJ}}$. As for the 
factor $z^{\beta}\cdot z^{2 - \beta}$ in the second term, it~is~$2(\alpha_R - \alpha_B)$~\cite{KTM}. 
For the third term  one has
just added an extra factor $z^{1/2}$.

In~\cite{ACKS} the value $\alpha_{SJ} = 0.5$ was used, while in
\cite{ShBopp} $\alpha_{SJ} = 0.9$. The values of $\epsilon, v_0, v_q$, 
and $v_{qq}$ in (\ref{t7}) were taken from \cite{ACKS, ShBopp}.
All other model parameters were also taken from \cite{KTM,Sh,ACKS, ShBopp}.

The probabilities of transition into the secondary baryon of  SJ without
valence quarks, $I_3$, SJ plus one valence quark, $I_2$, and
SJ plus a valence diquark, $I_1$,  follow
the simplest quark  combinatorials \cite{CS}. Assuming  the same strange
quark suppression in all the cases, we obtain, for the relative yields of different
baryons from SJ fragmentation without valence quarks:

\begin{equation}
\label{t8}
I_3 = 4L^3 : 4L^3 : 12L^2S : 3LS^2 : 3LS^2 : S^3,
\end{equation}

\noindent
for secondary $p$, $n$, $\Lambda + \Sigma$, $\Xi^0$, $\Xi^-$,  and
$\Omega$, respectively.

For $I_2$ we obtain

\begin{equation}
\label{t9}
I_{2u} = 3L^2 : L^2 : 4LS : S^2 : 0,
\end{equation}
and

\begin{equation}
\label{t10}
I_{2d} = L^2 : 3L^2 : 4LS : 0 : S^2,
\end{equation}
for secondary $p$, $n$, $\Lambda + \Sigma$, $\Xi^0$, and  $\Xi^-$.

For $I_1$  we get

\begin{equation}
\label{t11}
I_{1uu} = 2L : 0 : S,
\end{equation}
and

\begin{equation}
\label{t12}
I_{1ud} = L : L : S,
\end{equation}

\noindent
for secondary $p$, $n$, and $\Lambda + \Sigma$. The ratio $S/L$
determines the strange suppression factor and $2L + S$ = 1.
In the numerical calculations we  have used $S/L = 0.2$.

In agreement with  empirical rules we assume that
$\Sigma^+ + \Sigma^- = 0.6\cdot\Lambda$ \cite{appelsh} in Eqs.~(\ref{t8})-(\ref{t12}).
As customary $\Sigma^0$ are included into $\Lambda$. Note that the empirical rule
used in many experimental papers is not consistent with the simplest
quark statistics \cite{AKNS}.

The values of $v_0$, $v_q$, and $v_{qq}$ are  directly determined by the
corresponding coefficients of Eqs.~(\ref{t8})-(\ref{t12}), and by the
probabilities to fragment a $qqs$ system into $\Sigma^+ + \Sigma^-$ and $\Lambda$  (see above).

For incident $uu$ diquark and secondary $\Lambda$:

\begin{equation}
\label{t13}
v_0 = \frac{12}{1.6}S L^2 \;\;, v_q = \frac4{1.6}S L \;\;,
v_{qq} = \frac14 S \; ;
\end{equation}
for incident $ud$ diquark and secondary $\Lambda$:

\begin{equation}
\label{t14}
v_0 = \frac{12}{1.6}S L^2 \;\;, v_q = \frac4{1.6}S L \;\;,
v_{qq} = S \; .
\end{equation}

 As discussed above, although in the approach \cite{IOT,RV} the baryon consists of
three valence quarks together with SJ, which is
conserved during the interaction, it has been argued \cite{14r} that the annihilation cross section
contains a small  component, which is independent of $s$ and  makes $\alpha_{SJ}
\sim1$.  In any case, and with independence of the value of $\alpha_{SJ}$, the graph corresponds to an annihilation
contribution, and, therefore, it is weighted by a small
coefficient which will be denoted by $\varepsilon$. In our calculation we
 will use both values of $\alpha_{SJ}=0.5$ \cite{ACKS} and $\alpha_{SJ}=0.9$ \cite{ShBopp} and treat $\varepsilon$
as a free parameter.

\vskip 0.9 truecm
\noindent{\bf 3. COMPARISON WITH THE DATA}
\vskip 0.5 truecm

The mechanism of the baryon charge transfer via SJ without valence
quarks  was not accounted for in  papers \cite{KTM,2r,KTMS,Sh}.
As it was shown in \cite{ACKS,ShBopp} the data at comparatively low 
energies ($\sqrt{s} \sim 15 \div 40$~GeV) can be described with the 
values $\alpha_{SJ} = 0.5$ and $\varepsilon = 0.05$ (see \cite{ACKS}), 
or $\alpha_{SJ} = 0.9$ and $\varepsilon = 0.024$ \cite{ShBopp}.

The inclusive spectra of secondary $\Lambda$ produced in $\pi p$ 
collisions at energies 147 GeV/c \cite{Brick} and 250 GeV/c 
\cite{Ajinenko,Bogert} are shown in Fig.~1 together with the curves 
calculated in the QGSM at 250 GeV for both values of  $\alpha_{SJ}$.
The full line corresponds to $\alpha_{SJ} = 0.5$ \cite{ACKS}, and the 
dashed line to $\alpha_{SJ} = 0.9$ \cite{ShBopp}. The agreement of our 
results  with the existing experimental data is quite reasonable,
although one can see that at high values of $x_F>0.6$, the theoretical 
curves have remarkable different behavior. Unfortunately, due to the absence 
of the experimental data one cannot distinguish between the two values of  
$\alpha_{SJ}$.

The inclusive spectra of  $\bar \Lambda$ produced in $\pi p$ collisions  
at $E_{lab}=250 GeV/c$ \cite{ Ajinenko, Bogert} are shown in Fig.~2. The 
theoretical curves  do not depend on the values of $\alpha_{SJ}$.

The data on $y$-dependence of cross sections of $\Lambda$ production at 
energies 100, 175, and 360 Gev/c \cite{Biswas}, together with 250 GeV/c 
data \cite{ Bogert} are compared with QGSM calculations in Fig.~3.

The model calculations were made for energies 100 and 360 GeV/c and both 
values of $\alpha_{SJ}$: 1) dashed-dotted line corresponds to calculations for  
360 Gev/c and $\alpha_{SJ}=0.9$; 2) full line, for  360 Gev/c and 
$\alpha_{SJ}=0.5$; 3) 100 Gev/c and $\alpha_{SJ}=0.9$, and 4) 100 Gev/c 
and $\alpha_{SJ}=0.5$. Some disagreements of the calculated curves with 
the data are of the same order as the disagreement among experimental 
data.

In the same way, the data on $y$-dependence of cross sections of  
$\bar \Lambda$ production at energies 100, 175, and 360 Gev/c from 
Ref.~\cite{Biswas}, together with  250 GeV/c data from
Ref.~\cite{Bogert} are shown in Fig. 4.

The spectra of antibaryons, produced in $\pi^-p$ collisions, Figs.~2 and 
4, allow one to fix the fragmentation function of a quark into antibaryon.  
Since the  fragmentation functions into $\bar \Lambda$ do not depend on 
the SJ contribution, the $\bar \Lambda$ spectra are the same when 
calculated with different values of $\alpha_{SJ}$.

In Fig.~5 we show the data on the asymmetry of $\Lambda / \bar \Lambda$
produced in $\pi p$ interactions at 250 GeV/c \cite{alves} and 500 GeV/c  
\cite{ait1}. The asymmetry is defined as

\begin{equation}
\label{t15}
A(B/\bar{B}) = \frac{N_B - N_{\bar{B}}}{N_B + N_{\bar{B}}},
\end{equation}
for each $x_F$ bin.

The theoretical curves are calculated for 500 GeV/c with $\alpha_{SJ}=0.5$ 
(full line) and $\alpha_{SJ}=0.9$ (dashed line). Both calculations are in 
reasonable agreement with the existing data.

The data from reference \cite{alves}, measured at rather large interval 
of $x_F$, were nott available during the preparation of this paper \cite{ACKS}.

As we can see from the present calculations, the existing data on strange 
baryon  production are reasonably compatible with both values of 
$\alpha_{SJ}$.  However, they show remarkably different behaviors for 
larger values of $x_F$ in the forward hemisphere.

\vskip 0.9 truecm
\noindent{\bf 5. CONCLUSIONS}
\vskip 0.5 truecm

We have  shown that experimental data on high-energy $\Lambda$ production
support the possibility of baryon charge transfer over large
rapidity distances.   The asymmetry is provided by
string junction diffusion through baryon charge transfer.

The production of net baryons in $\pi p$ interactions
in the projectile hemisphere provides good evidence for such a mechanism.

As for the values of the parameters $\alpha_{SJ}$ and $\varepsilon$ which
govern the baryon charge  transfer, we have seen that the data on strange 
baryon  production at comparative low energies favor both the values 
$\alpha_{SJ} = 0.5$ and $\alpha_{SJ} =0.9$.

The importance of understanding the dynamics of the transfer of baryon charge over large rapidity
distances stresses the necesity of good experimental data in meson and baryon
collisions with nucleons and nucleus, as well as in nucleus-nucleus interactions for different centralities.

For the $\Lambda$ production on  $\pi$ beams the disagreement
between the theoretical calculations for the two different values of
$\alpha_{SJ}=0.5, 0.9$, are within the experimental uncertainty.\\

\vskip 0.5 truecm
\noindent {\bf Acknowledgments}
\vskip 0.2 truecm

 The authors are thankful to A. Capella, C. Pajares,  A. B. Kaidalov, and  E. G. Ferreiro
for useful discussions.

This work was partially financed by CICYT of Spain through contract FPA2002-01161, and by Xunta de Galicia
through contract PGIDIT03PXIC20612PN.

G. H. Arakelyan and C.Merino were also supported by NATO grant CLG 980335 and
Yu. M. Shabelski by grants NATO PDD (CP) PST.CLG 980287, RCGSS-1124.2003.2.

\newpage

\vskip 1.2cm
{\bf Figure Captions}
\vskip 0.2 truecm

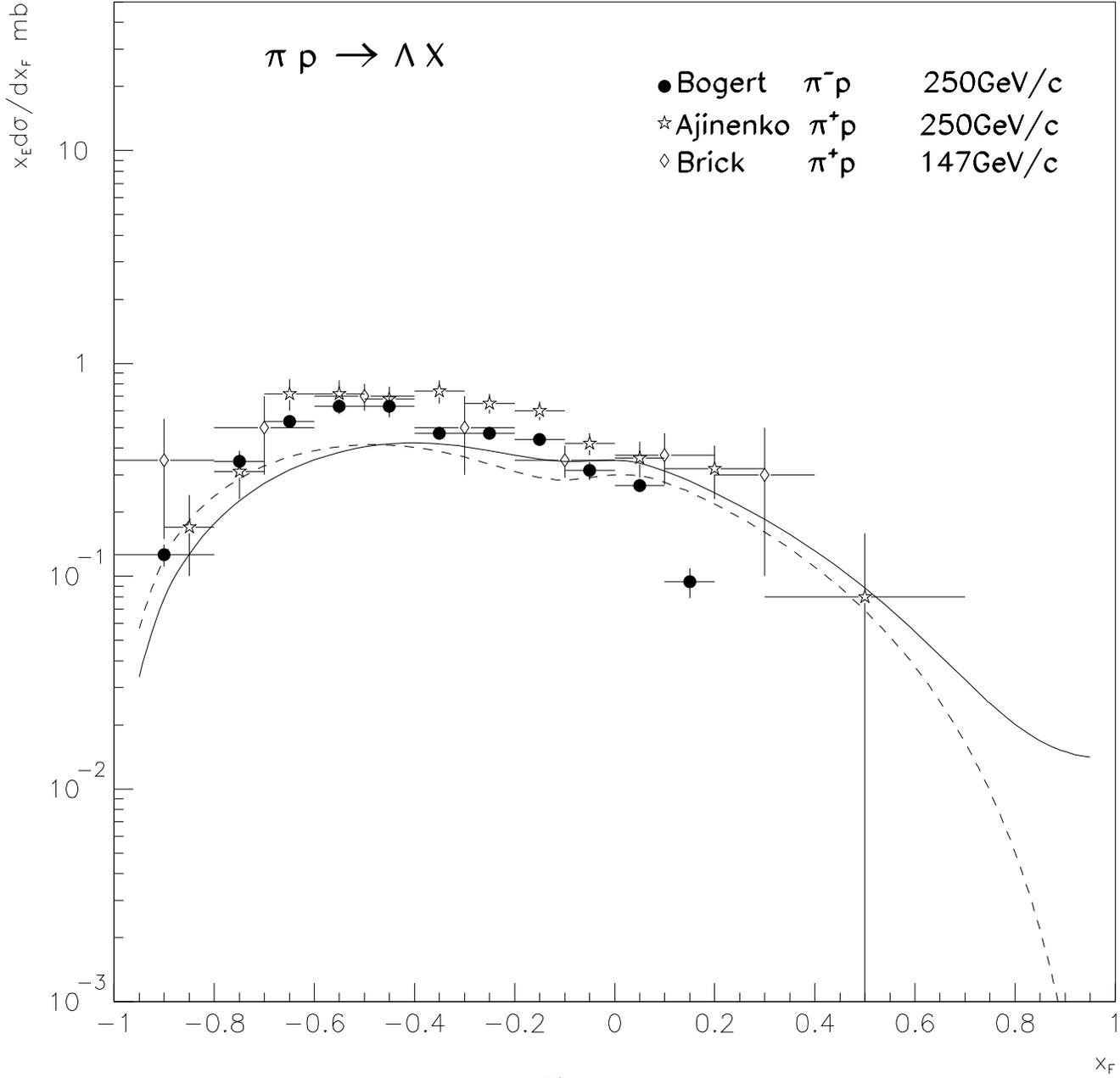
Fig. 1.  The experimental data on $x_F$ spectra of secondary $\Lambda$ in $\pi^{\pm} p$ collisions
at 147 GeV/c \cite{Brick} and 250 GeV/c \cite{ Ajinenko, Bogert}, together with  QGMS description. The full line corresponds to
calculations  with $\alpha_{SJ}=0.5$, while the dashed line corresponds to $\alpha_{SJ} =0.9$.

Fig. 2.  The $x_F$ spectra of secondary $\bar\Lambda$ in $\pi p$ collisions. Experimental data
at 250 GeV/c \cite{Ajinenko, Bogert} and QGSM description.

Fig. 3.  The $y$-dependence of spectra of secondary $\Lambda$ in $\pi p$ 
collisions. Experimental data at 100, 175, and 360 GeV/c \cite{Biswas}, 
and 250 GeV/c\cite{Bogert}. The QGSM description at $E=100$ and 360 Gev/c 
for both values of $\alpha_{SJ}$.

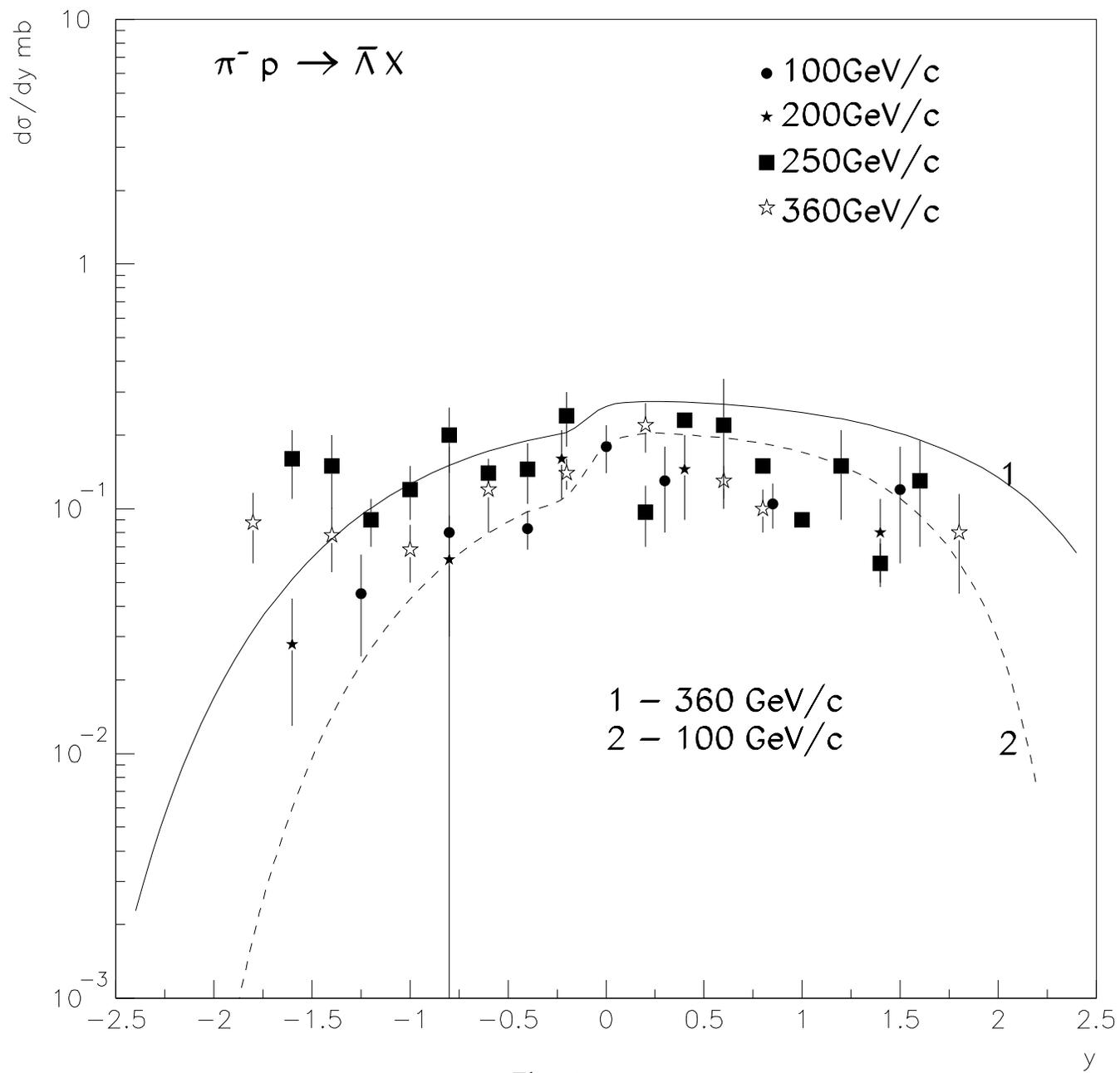
Fig. 4.  The $y$-dependence of spectra of secondary $\bar \Lambda$ in 
$\pi p$ collisions. Experimental data at 100, 175,  360 GeV/c 
\cite{Biswas}, and 250 GeV/c \cite{Bogert}. The QGSM description at 
$E=100$ Gev/c (dashed line) and 360 Gev/c (full line).

Fig. 5.  The QGSM description of the asymmetry $\Lambda/\bar{\Lambda}$ in $\pi^-p$ collisions
at 500 GeV/c \cite{ait1} and 250 GeV/c \cite{alves}. The curves were calculated for 500 GeV/c with
 $\alpha_{SJ}=0.5$ (full line) and $\alpha_{SJ}=0.9$ (dashed line).

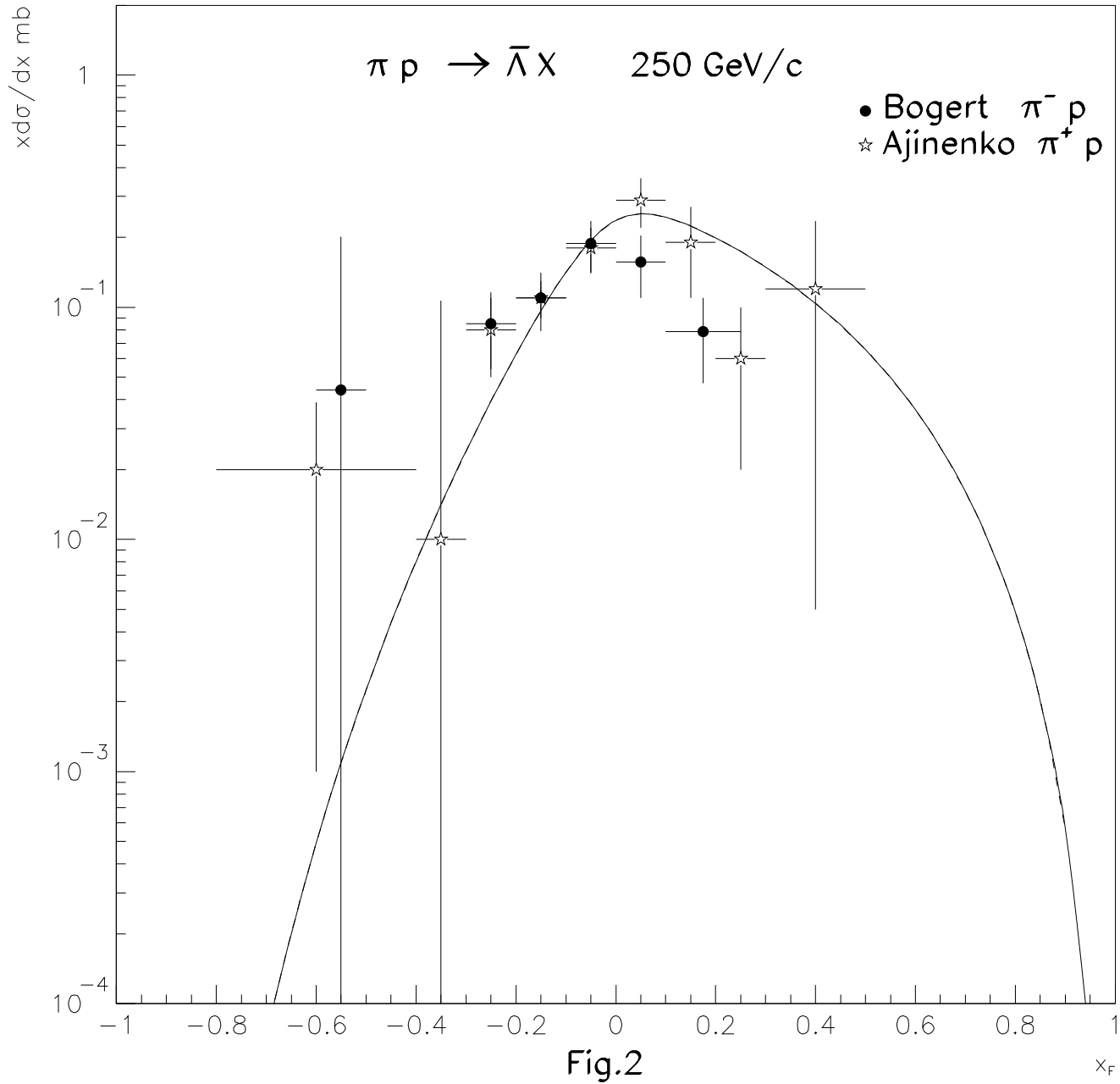
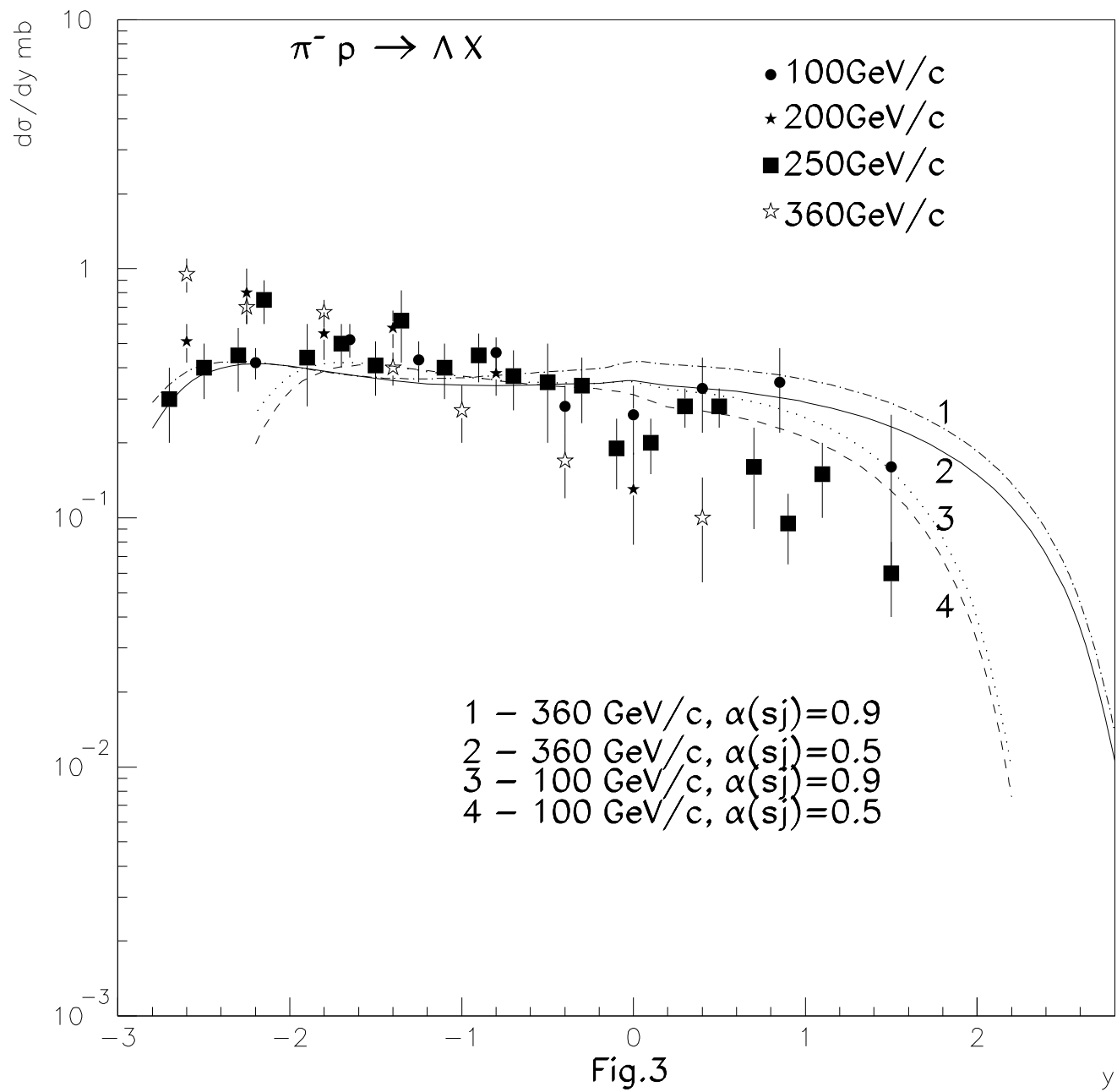
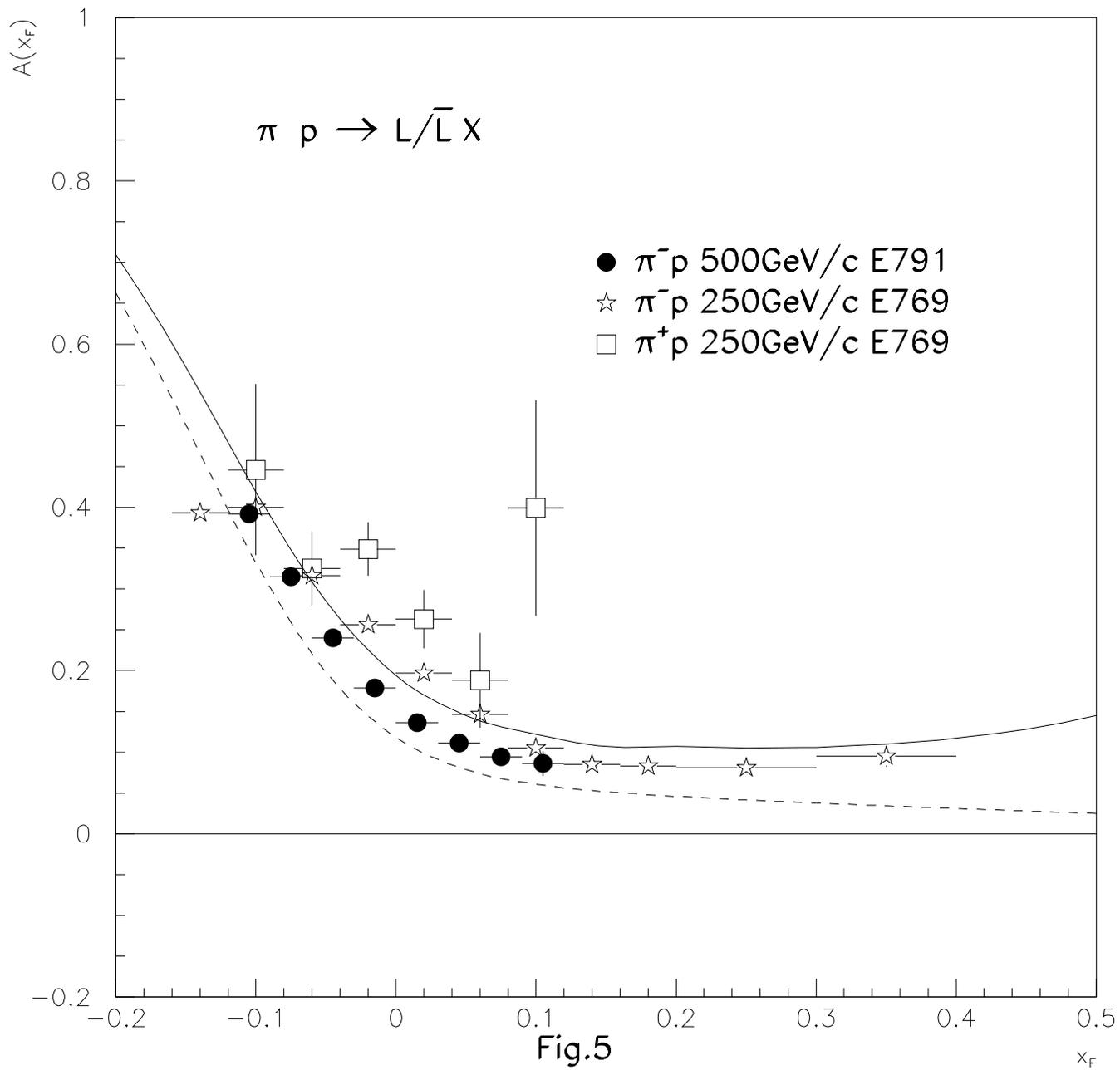
\end{document}